\documentclass[12pt]{article}
\usepackage{amsmath,amssymb,exscale,cancel}
\usepackage{slashed}
\allowdisplaybreaks
\usepackage{graphicx}
\usepackage{graphicx}
\usepackage{epsfig}
\usepackage{multicol}
\usepackage{color}
\usepackage{mathrsfs}
\usepackage{blindtext}
 \usepackage{fancyhdr}
\usepackage{hyperref}
\usepackage{cite}
\usepackage{mathtools}

\usepackage{floatrow}

\usepackage{graphicx}
\usepackage{sidecap}

\usepackage[latin1]{inputenc} 

\usepackage{multicol}  
 
 \usepackage{titlesec}
 
 \usepackage{rotating,slashed,xcolor,amsfonts,expdlist,charter}

\numberwithin{equation}{section}

\usepackage{xcolor}
\usepackage{sectsty}

\usepackage{mdframed}
\usepackage{titletoc}

\numberwithin{equation}{section}

\usepackage{xcolor}
\usepackage{sectsty}

\usepackage{mdframed}
\usepackage{titletoc}

\definecolor{secnum}{RGB}{13,151,225}
\definecolor{ptcbackground}{RGB}{212,237,252}
\definecolor{ptctitle}{RGB}{0,177,235}

\titlecontents{lsection}
  [5.8em]{\sffamily}
  {\color{secnum}\contentslabel{2.3em}\normalcolor}{}
  {\titlerule*[1000pc]{.}\contentspage\\\hspace*{-5.8em}\vspace*{5pt}%
    \color{white}\rule{\dimexpr\textwidth-15.5pt\relax}{1pt}}

\usepackage{hyperref}
\hypersetup{colorlinks,bookmarksopen,bookmarksnumbered,citecolor=rossos,
linkcolor=redy,pdfstartview=FitH,urlcolor=green-go}
\usepackage{slashed}

\definecolor{blus}{cmyk}{1,0.9,0,0.1}
\definecolor{verdes}{cmyk}{0.99,0,0.59,0.65}
\definecolor{rossos}{cmyk}{0,1,1,0.55}
\definecolor{redy}{cmyk}{0,1,1,0.7}
\definecolor{greeny}{cmyk}{0.99,0,0.59,0.98}
\definecolor{green-go}{cmyk}{0.79,0,0.59,0.5}

\usepackage{titlesec}

\newcommand{\beq}{\begin{equation}}
\newcommand{\eeq}{\end{equation}}

\def\hhref#1{\href{http://arxiv.org/abs/#1}{arXiv:#1}} 

\newcommand{\tmtextbf}[1]{{\bfseries{#1}}}
\newcommand{\tmtextrm}[1]{{\rmfamily{#1}}}

\def\be{\begin{equation}}
\def\ee{\end{equation}}
\def\ba{\begin{array} }
\newcommand{\Tr}{\,{\rm Tr}}
\def\bac{\begin{array} {c}}
\def\bacc{\begin{array} {cc}}
\def\baccc{\begin{array} {ccc}}
\def\bacccc{\begin{array} {cccc}}
\def\ea{\end{array}}
\def\bea{\begin{eqnarray}}
\def\eea{\end{eqnarray}}

\definecolor{red}{rgb}{1,0,0}

\def\psl{\hbox{\hbox{${p}$}}\kern-1.9mm{\hbox{${/}$}}}
\def\dsl{\hbox{\hbox{${\partial}$}}\kern-2.2mm{\hbox{${/}$}}}
\def\Dsl{\hbox{\hbox{${D}$}}\kern-2.6mm{\hbox{${/}$}}}

\newcommand{\gappeq}{{\rlap{{\raise}.5ex\text{\ensuremath{>}}}{{\lower}.5ex\text{\ensuremath{\sim}}}}}
\newcommand{\lappeq}{{\rlap{{\raise}.5ex\text{\ensuremath{<}}}{{\lower}.5ex\text{\ensuremath{\sim}}}}}
\newcommand{\I}{\tmtextrm{1{\kern}-.24em l}}

\begin{document}
\topmargin -1.0cm
\oddsidemargin 0.9cm
\evensidemargin -0.5cm

{\vspace{-1cm}}
\begin{center}

\vspace{-1cm}

 {\tmtextbf{ 
 \hspace{-1.2cm}   
{\LARGE  \color{rossos}Thermal Gauge Theory for a Rotating Plasma}
 \hspace{-1.6cm}}} {\vspace{.5cm}}\\

\vspace{1.3cm}

{\large{\bf  Alberto Salvio }}

{\em  
\vspace{.4cm}
 Physics Department, University of Rome Tor Vergata, \\ 
via della Ricerca Scientifica, I-00133 Rome, Italy\\

\vspace{0.6cm}

I. N. F. N. -  Rome Tor Vergata,\\
via della Ricerca Scientifica, I-00133 Rome, Italy\\

  \vspace{0.5cm}

}
\vspace{1.5cm}
\end{center}

\noindent ---------------------------------------------------------------------------------------------------------------------------------
\begin{center}
{\bf \large Abstract}
\end{center}
\noindent   This paper provides a systematic and complete study of thermal gauge theory for generic equilibrium density matrices, which feature arbitrary values not only of temperature and chemical potentials, but also of average angular momentum. This work extends previous studies, which focused on pure scalar-fermion theories, to all gauge theories coupled to an arbitrary matter sector. Path-integral methods are developed to study ensemble averages and thermal Green's functions of general operators, with an arbitrary number of points, in all interacting gauge theories. These methods cover both the real-time and imaginary-time formalisms. Generalized Kubo-Martin-Schwinger (KMS) conditions are obtained both in coordinate and in momentum space for operators in general representations of the Lorentz and internal symmetry group. This allows us to obtain all thermal propagators including those of gauge fields and Faddeev-Popov ghosts. By analyzing all interactions in detail, it is shown that, in perturbation theory, only the propagators are affected by the average angular momentum and the chemical potentials, the vertices remain unmodified. The paper presents fully model-independent results and can, therefore, be applied to any specific thermal field theory.

\vspace{0.7cm}

\noindent---------------------------------------------------------------------------------------------------------------------------------

\newpage

\tableofcontents

\noindent --------------------------------------------------------------------------------------------------------------------------------

\vspace{0.2cm}

\section{Introduction}\label{intro}

Thermal field theory (TFT) allows us to describe physical systems with a huge number of particles, where relativistic and/or quantum effects are important. Applications of TFT include, among others, the description of the early universe after the reheating period and the investigation of compact astrophysical objects such as neutron stars, (primordial or astrophysical) black holes, and more exotic compact objects. Moreover, TFT is now the standard tool to study particle-physics processes (particle decay, scattering, and production processes) in a plasma, as well as phase transitions in a relativistic quantum setup (see~\cite{Bellac:2011kqa,Nair:2005iw} for textbooks,~\cite{Landsman:1986uw,Quiros:1994dr,Laine:2016hma} for monographs, and~\cite{Salvio:2024upo} for an introduction from first principles).

At thermodynamic equilibrium, the density matrix $\rho$, which is the key input in TFT, can be expressed in terms of all conserved quantities: the Hamiltonian $H$; the linear and angular momentum, $\vec P$ and $\vec J$, respectively; and all conserved charges $Q^a$ associated with an internal symmetry group $\mathcal{G}$~\cite{LandauLifshitz}. Choosing the reference frame appropriately (see Ref.~\cite{Salvio:2025rma} for details), the most general equilibrium density matrix can always be written as
\be
\rho = \frac{e^{-\beta (H - \vec\Omega \cdot \vec J - \mu_a Q^a)}}{Z},
\label{rhoRest}
\ee
where $Z$ is the partition function, $\beta \equiv 1/T$ is the inverse temperature, and $\mu_a$ is the chemical potential associated with $Q^a$. Moreover, $\vec\Omega$ is a thermodynamic quantity associated with the average angular momentum of the system. Sometimes $\vec\tau \equiv -\beta \vec\Omega$ is referred to as the thermal vorticity.

Two previous papers~\cite{Salvio:2025rma,Salvio:2025ggj} initiated, in the case of pure scalar--fermion theories, a systematic study of (generically interacting) TFT for the most general equilibrium density matrix, including not only $T$ and all $\mu_a$, but also a non-vanishing value of the average angular momentum\footnote{See also e.g.~\cite{Zubarev:1979afm,Weert,Becattini:2012tc} for previous studies of the density matrix in~(\ref{rhoRest}), and Refs.~\cite{Vilenkin:1980zv,Buzzegoli:2017cqy,Prokhorov:2019yft,Becattini:2020qol,Kuboniwa:2025vpg,Siri:2025kdw,Ambrus:2015lfr,Chernodub:2016kxh,Vilenkin:1978hb,Vilenkin:1979ui,Buzzegoli:2020ycf,Palermo:2021hlf,Ambrus:2021eod,Palermo:2023cup,Castano-Yepes:2025zae,Zhu:2025pxh} for previous studies of specific scalar and/or fermion TFTs in the presence of a rotating plasma.}, corresponding to $\vec\Omega \neq 0$.

The present work extends the analysis of Refs.~\cite{Salvio:2025rma,Salvio:2025ggj} to all classes of field theories, including gauge theories, while keeping the most general equilibrium density matrix, with arbitrary values of the average angular momentum, $T$, and all possible $\mu_a$ (some of the $\mu_a$ may correspond to global charges, while others correspond to gauge charges). The inclusion of spin-1 particles and gauge interactions is essential for realistic applications of TFT, even in the presence of rotation and chemical potentials. For example, color chemical potentials in QCD are introduced to study color superconductivity (see~\cite{Alford:2007xm,Schmitt:2025cqi} for reviews), which may occur in the interior of (rotating) compact stars such as neutron stars, e.g.~pulsars. Moreover, extensions of the Standard Model in which its internal symmetries are extended and/or gauged are well motivated: gauge-group extensions are suggested, for example, by grand unification and the possible existence of gauge dark sectors, while the gauging of existing global symmetries is motivated, for instance, by some approaches to quantum gravity. Last but not least, photons may be in thermal equilibrium with, and/or produced by, rotating plasmas with sizable values of the temperature and/or chemical potentials.

In this paper, a model-independent approach is adopted, which provides ready-to-use formul\ae~for perturbative and non-perturbative calculations once a specific gauge TFT for a rotating plasma is chosen\footnote{See e.g.~Refs.~\cite{Vilenkin:1980zv,Buzzegoli:2020ycf,Braguta:2021jgn,Braguta:2024zpi,Castano-Yepes:2025zae} for previous studies of specific gauge TFTs in the presence of a rotating plasma.}.

Because of complications due to the fact that gauge theories are  constrained systems, functional and path-integral methods are employed here. Moreover, in order to make the final results applicable to both perturbative and non-perturbative calculations, a formalism that includes both the real-time and imaginary-time approaches as special cases is used.

The path-integral approach is complemented by a generalized form of the Kubo--Martin--Schwinger (KMS) condition. In~\cite{KMS}, Kubo, Martin, and Schwinger proposed an exact (non-perturbative) relation between the two non-time-ordered 2-point functions of a pair of field operators, $\mathcal{O}_1$ and $\mathcal{O}_2$, in finite-temperature field theory. Here, this condition is extended to include generic values of both $\vec\Omega$ and the $\mu_a$, for arbitrary Lorentz and internal-symmetry representations of $\mathcal{O}_1$ and $\mathcal{O}_2$ (generalized KMS condition). Moreover, some analyticity properties of the corresponding 2-point functions are investigated.

The KMS condition provides a quantum version of the fluctuation-dissipation theorem~\cite{Pappalardi:2021ahe}.
Another important application of the KMS condition\footnote{Yet another application of the KMS condition furnishes a simple relation between absorption and production rates of particles coupled to the plasma (see, e.g., Ref.~\cite{Altherr:1993tn}).} is the computation of thermal propagators. In this work, the generalized KMS conditions are used to compute all thermal propagators in a generic gauge TFT in the presence of arbitrary values of $\vec\Omega$ and the $\mu_a$.

Furthermore, the paper provides a model-independent analysis of the effects of $\vec\Omega$ on interactions in gauge TFTs, and how $\vec\Omega$ affects perturbation theory (when the perturbative expansion is valid).

The paper is organized as follows.
\begin{itemize}
\item In the next section, preliminary remarks on free spin-1 particles are presented to illustrate how ensemble averages can be determined in this simple case. In the same section, it is also explained in more detail why the structure of gauge theories in practice demands the use of functional and path-integral methods.
\item In Sec.~\ref{Classical gauge theory} the classical gauge theory is discussed, in particular in an axial gauge, which is convenient for performing the subsequent quantization for general equilibrium density matrices with average angular momentum (as well as $T$ and the $\mu_a$).
\item Sec.~\ref{Quantization and general path-integral formula} contains the quantization procedure. After quantization in the axial gauge, path-integral expressions for the partition function and the thermal Green's functions are derived using a general gauge fixing, while keeping the density matrix in its most general equilibrium form.
\item The generalized KMS conditions are derived in Sec.~\ref{Generalized Kubo-Martin-Schwinger condition} in both coordinate and momentum space.
\item All thermal propagators of a gauge TFT with general values of $T$, $\mu_a$, and $\vec\Omega$ are computed in Sec.~\ref{propGauge}.
\item The effects of $\vec\Omega$ on interactions in gauge TFTs, and its role in perturbation theory, are discussed in Sec.~\ref{Interactions}.
\item Sec.~\ref{Conclusions} provides a detailed summary of the main original results and the final conclusions.
\end{itemize}

\section{Preliminary remarks on free spin-1 particles}\label{Preliminary remarks on free spin-1 particles}
  
  Before proceeding to the discussion at the fully interacting level, let us make some remarks on free spin-1 particles. 
  
  It is easy to extend the results of Sec.~3.1 of~\cite{Salvio:2025rma} for free spin-0 particles to free spin-1 particles.  Setting the third axis of the reference frame along the direction of $\vec\Omega$, one can write spin-1 one-particle states in a basis of eigenstates of $H$, the third components of the linear and angular momentum, $P^3$ and $J_3$ respectively, and\footnote{In this work repeated indices understand a summation (unless otherwise stated).} $\vec J\cdot \vec P/|\vec P| \equiv J_kP^k/|\vec P|$, with the latter operator having two (three) eigenvalues for massless (massive) spin-1 particles;
   this, of course, corresponds to the respective two (three) degrees of freedom of a massless (massive) spin-1 particle. Adapting the discussion of Sec.~3.1 of~\cite{Salvio:2025rma}, one then finds that the partition function of a free spin-1 particle is the same as  the partition function of a free spin-0 particle with the same mass and in the same irreducible representation of $\mathcal{G}$ raised to a power equal to the number of degrees of freedom.  Then the ensemble averages of $H$, $\vec J$ and $Q^a$ for a free spin-1 particle are equal to  the corresponding quantities for a free spin-0 particle, but multiplied by the number of degrees of freedom. 

Also, enclosing the rotating system in a cylinder of radius $R$, the convergence of the averages for spin-1 particles, just like the convergence of the  averages for spin-0 particles,
requires the bound $\Omega\equiv|\vec\Omega|<1/R$, so that 
\be v\equiv \Omega R\in[0,1), \label{vrange} \ee
which agrees with the fact that the particles in the rotating plasma must not exceed the speed of light, identifying $\vec \Omega$ with the angular-velocity vector of the rotating plasma (see Sec.~\ref{Quantization and general path-integral formula}).  When $\Omega$ is not zero one needs to keep $R$ finite, as a consequence of~\eqref{vrange}~\cite{Vilenkin:1980zv,Ambrus:2015lfr,Ebihara:2016fwa,Chernodub:2016kxh}.

  As an example of how the energy and angular-momentum densities depend on $\Omega$ for a free spin-1 particle one can look at  Fig.~2 of~\cite{Salvio:2025rma} multiplying the vertical axis by the number of degrees of freedom of the thermalized  spin-1 particle in question (see~\cite{dataset,datasetf} for the datasets of Refs.~\cite{Salvio:2025rma,Salvio:2025ggj}). 
    
    One might then think to go ahead and compute all thermal propagators of a gauge theory (including those of the Faddeev-Popov ghosts) by using the operator formalism with creation and annihilation operators, as was done for scalars and fermions in~\cite{Salvio:2025rma,Salvio:2025ggj}.
  However, because gauge theories are constrained systems,
  it is prohibitively difficult to do so in practice.  
  For this reason here another road,  based on functional methods and path integrals, will be followed. This, as we shall see, in particular avoids the difficulties due to the presence   of secondary constraints provided by the field equations for the temporal components of the gauge fields. At the same time, this will also allow us to have a recipe to deal with interactions in general (thermal) gauge theories.

 \section{Classical gauge theory}\label{Classical gauge theory}
  
 Let us first start from the classical  theory and proceed afterwards to quantization in Sec.~\ref{Quantization and general path-integral formula}. After all, the most famous gauge theory, electrodynamics, was first known as a classical theory and was quantized only afterwards.
 
  We consider, however,  a general Yang-Mills (not necessarily Abelian) theory, where the gauge group $G$ is a Lie group  with generators $\mathcal{T}^a$, with $a=1, ... , \dim G$. The global subgroup of $G$ can be viewed as a subgroup of the full global group $\mathcal{G}$. We choose the $\mathcal{T}^a$ such that the real constants $f^{abc}$ (the structure constants), defined by
\be [\mathcal{T}^a,\mathcal{T}^b] = i f^{abc} \mathcal{T}^c, \ee
are totally antisymmetric.
  The field strength $F^a_{\mu\nu}$ is given in terms of the gauge fields $A^a_\mu$ by
\be F^a_{\mu\nu} = \partial_\mu A^a_\nu-\partial_\nu A^a_\mu -g f^{bca} A^b_\mu A^c_\nu,\ee
where $g$ represents the gauge couplings (here a compact notation is used and an index on $g$ is implied, so also the cases with more than one gauge coupling are covered).

We start from the gauge-invariant  classical Lagrangian given by
\be \mathscr{L}= -\frac14 F^a_{\mu\nu}F^{a\mu\nu} + \mathscr{L}_M(\Theta,D\Theta), \ee
where $\mathscr{L}_M$ represents the matter Lagrangian, which we assume here to be an ordinary function of only the matter fields $\Theta$ (which can be scalars and/or fermions) and their covariant derivatives $D_\mu\Theta$, given by
\be D_\mu\Theta=(\partial_\mu+i g\mathcal{T}^a A^a_\mu)\Theta.\ee 

To exploit the cylindrical symmetry induced by $\vec\Omega$ let us choose the axial gauge 
\be A_3^a = 0. \ee
Then the canonical variables of the gauge sector are the $A_i^a$, with $i=1,2$, and the corresponding conjugate momenta are
\be \Pi_a^{i} \equiv \frac{\partial\mathscr{L}}{\partial \dot  A_i^a}  = F^{a}_{0i}, \ee
while $A_0^a$ is not an independent variable, but is a functional of the other variables: such functional is obtained by solving the field equations of $A_0^a$ itself, which represent 
secondary constraints~\cite{Weinberg2}. 

\section{Quantization and general path-integral formula}\label{Quantization and general path-integral formula}

  One can now express a generic thermal Green's function of the form\footnote{For a generic operator $\mathcal{F}$ the ensemble average is 
$\langle \mathcal{F}\rangle =  \Tr(\rho \mathcal{F})$.
}
  \be  \langle {\cal T}\mathcal{O}_1(x_1) ...  \mathcal{O}_n(x_n) \rangle= \Tr (\rho \mathcal{T} \mathcal{O}_1(x_1) ...  \mathcal{O}_n(x_n)), \label{GreenO} \ee
   where now the $\mathcal{O}_i$ are generic local 
   operators involving gauge fields, in terms of a path integral over  $A_i^a$  and $\Pi_a^{i}$.

   Combining the derivation of such path integral in purely scalar theories, but with a general equilibrium density matrix, 
   given in~\cite{Salvio:2025rma},   with that in a non-statistical gauge theory provided in~\cite{Weinberg2}, we find
\bea   \hspace{-0.7cm}\langle {\cal T}\mathcal{O}_1(x_1) ...  \mathcal{O}_n(x_n) \rangle =\frac1{Z}\int \delta A_1\delta A_2\, \delta \Pi^1\delta \Pi^2  \, O^\omega_1(x_1^\omega) ... O_n^\omega(x_n^\omega)\nonumber \\  \exp\left(i \int_C d^4x \left(\dot A_i(x)\Pi^i(x) - \mathcal{H}^\omega_c(A(x), \Pi(x))\right)\right).\label{PIgauge}\hspace{-0.3cm}\eea
   Some definitions and clarifications are now in order to understand this result.
   \begin{itemize}
\item First, the operators appearing in~(\ref{PIgauge}) are defined by $O^\omega_i(x_i^\omega)\equiv \mathcal{D}^{-1}_i(t_i\vec\Omega)O_i(x_i^\omega)$. Also,  $x^\omega\equiv \{t, R(t\vec\Omega)\vec x\}$ and $\mathcal{D}_i(t_i\vec\Omega)$  implement the rotation of angle $t_i\Omega$ around $\vec\Omega$ on the spacetime coordinates and in the Lorentz representation of $O_i(x_i)$, respectively. The $c$-number fields $O_i(x_i)$ are  obtained by substituting (in the field operators $\mathcal{O}_i(x_i)$) the canonical quantum variables with their eigenvalues after putting all canonical ``coordinates" (the $A_i^a$) on the right of all canonical ``momenta" (the $\Pi^i_a$). 
 The matrices $\mathcal{D}_i^{-1}(t_i\vec\Omega)$ appear because in the derivation one defines  new operators $\mathcal{\tilde O}_i$, which are related to the $\mathcal{O}_i$ by
  \be \mathcal{\tilde O}_i(t,\vec x) \equiv  \exp(i(H-\vec\Omega\cdot \vec J\,)t)\,\mathcal{O}_i(0,\vec x)\,\exp(-i(H-\vec\Omega\cdot \vec J\,)t) =  \mathcal{D}_i(t\vec\Omega)\mathcal{O}_i(t, R^{-1}(t\vec\Omega)\vec x), \label{tildeO}\ee
(here the repeated index $i$ does not understand a summation)  and the relation in~(\ref{tildeO}) can be inverted\footnote{The inversion can be performed as follows. Consider first
 \be \exp(-i\vec\Omega\cdot \vec J\, t')\,\mathcal{\tilde O}_i(t,\vec{x})\,\exp(i\vec\Omega\cdot \vec J\,t') = \mathcal{D}_i((t+t')\vec\Omega)\mathcal{O}_i(t, R^{-1}((t+t')\vec\Omega)\vec x),\ee
 where $t'$ is an arbitrary time. On the other hand, one also has
 \be \exp(-i\vec\Omega\cdot \vec J\, t')\,\mathcal{\tilde O}_i(t,\vec{x})\,\exp(i\vec\Omega\cdot \vec J\,t')  = \mathcal{D}_i(t'\vec\Omega) \mathcal{\tilde O}_i(t, R^{-1}(t'\vec\Omega)\vec x), \ee
 so, setting $t'=-t$, one finds~(\ref{InTr}).}  to obtain
\be \mathcal{O}_i(x_i)= \mathcal{D}_i^{-1}(t_i\vec\Omega) \mathcal{\tilde O}_i(x_i^\omega)\equiv O^\omega_i(x_i^\omega). \label{InTr}\ee
  \item In the argument of the exponential in the path integral, while the space integral has  no restriction,
 the integral over $t$ is performed on a contour $C$ in the complex $t$ plane that connects an arbitrary time  $t_0$ with $t_0-i\beta$ and contains the time components $x_1^0, ... , x_n^0$ of   $x_1, ... , x_n$. The arbitrariness of $t_0$ allows us to use the real- or the imaginary-time formalism by choosing $C$ appropriately.
\item The path integration is on all gauge-field configurations $A$ satisfying  the twisted periodicity condition:
  \be e^{\beta\mu_a f^a}A(t_0,\vec{x}) = A(t_0-i\beta,\vec{x}),\label{PerConA}\ee
  where the $f^a$ are the generators of $\mathcal{G}$ in the  representation of the gauge fields (the adjoint representation), which acts on the Lie-group index.  
  \item The integration measures on the $A_i^a$ and the $\Pi^i_a$ are
  \be \delta A_i = \prod_{x, a} dA^a_i(x), \quad \delta \Pi^i= \prod_{x, a} d(\Pi^i_a(x)/(2\pi)).  
 \ee
\item The function $\mathcal{H}^\omega_c$ is, in the classical limit, the full classical Hamiltonian density, including the effect of rotation. The defining property of this quantity is
  \be \int d^3x \, \mathcal{H}^\omega_c = H_c- \vec\Omega\cdot \vec J_c\, , \label{callHc}\ee  
  where $H_c$ and $\vec J_c$ are 
\be H_c(A, \Pi)\equiv \frac{\langle A|H|\Pi\rangle}{\langle A|\Pi\rangle}, \qquad \vec J_c(A, \Pi)\equiv \frac{\langle A|\vec J |\Pi\rangle}{\langle A|\Pi\rangle}  \label{HJc}  \ee 
and $|A\rangle$ and $|\Pi\rangle$ are the  eigenstates of the canonical  quantum variables (the $A_i^a$ and the $\Pi^i_a$, respectively). The quantum Hamiltonian $H$ is chosen in a way that   $\mathcal{H}_c$, which satisfies by definition $H_c =\int d^3 x\mathcal{H}_c$, reproduces  the known expression for the classical Hamiltonian density of the Yang-Mills theory in the absence of rotation: e.g.~in the axial gauge we are using      
\be\mathcal{H}_c= \Pi_a^{i} \left(\partial_i A_0^a+gf^{bca} A^b_0 A^c_i\right)+\frac{1}{2}\Pi_a^{i} \Pi_a^{i}  +\frac14 F^a_{ij}F^{a}_{ij}+\frac12 \partial_3 A^a_i\partial_3 A^a_i-\frac12 \partial_3 A^a_0\partial_3 A^a_0,\label{Hcgauge} \ee
with $A_0^a$ set equal to the solution of its  field equations. Then, $A_0^a$ generically turns out to be a complicated (even non-local) functional of the canonical  variables. This problem will be solved by introducing a further path integration.
\item Similarly, the quantum angular momentum $\vec J$ is chosen in a way that $\vec J_c$ reproduces  the known expression for the classical angular momentum of the Yang-Mills theory:  the components $J_{ck}$, with $k=1,2,3$, of $\vec J_c$ are given by
\be J_{ck} = \epsilon_{lmk} \int d^3 x\, (\Pi^p_a\, x^m\partial_l A_p^a- A^{a}_m \Pi^l_a ), \label{Jcc}\ee 
where $\epsilon_{lmk}$ is the totally antisymmetric Levi-Civita symbol (with $\epsilon_{123}=1$).
This formula holds beyond the axial gauge $A_3^a=0$ (if $A_3^a\neq0$ one defines $\Pi^3_a \equiv \frac{\partial\mathscr{L}}{\partial \dot  A_3^a}  = F^{a}_{03}$) and  can also be written in a vector notation:
\be \vec J_c = \int d^3x \,(-\Pi^k_a\, \vec x \times \vec\nabla A_k^a + \vec A^a \times \vec \Pi_a), \label{Jcvector} \ee 
where $\vec A^a$ and $\vec \Pi_a$ are the three-vectors with components $A^a_k$ and $\Pi_a^k$, respectively.
The first term in~(\ref{Jcvector}) represents the orbital angular momentum contribution, while the second one is the spin contribution.
\end{itemize}

Using the cylindrical coordinates, 
\be x^1 = r\cos\phi, \qquad x^2 = r\sin\phi, \qquad x^3 = z\label{CyCoo}, \ee
one can now combine~(\ref{Hcgauge}) and~(\ref{Jcc}) to find (in the axial gauge)
\bea\mathcal{H}_c^\omega&=& \Pi_a^{i}\left[\partial_i A_0^a+gf^{bca} A^b_0 A^c_i +\Omega (\partial_\phi A_i^a+\epsilon_{ij} A^{a}_j)\right] \nonumber \\&&+\frac{1}{2}\Pi_a^{i} \Pi_a^{i}  +\frac14 F^a_{ij}F^{a}_{ij}+\frac12 \partial_3 A^a_i\partial_3 A^a_i-\frac12 \partial_3 A^a_0\partial_3 A^a_0,
\label{HcOmegagauge}\eea
 where $\epsilon_{ij}\equiv\epsilon_{ij3}$.

 The partition function that appears in the denominator in~(\ref{PIgauge}) can be expressed as
  \be \boxed{Z =\int \delta A_1\delta A_2 \, \delta \Pi^1\delta \Pi^2  \,  \exp\left(i \int_C d^4x \left(\dot A_i(x)\Pi^i(x) - \mathcal{H}^\omega_c(A(x), \Pi(x))\right)\right),} \label{GenPartGauge}  \ee
  which is nothing but the path integral in~(\ref{PIgauge}) without $O^\omega_1(x_1^\omega) ... O_n^\omega(x_n^\omega)$. Recall that $Z$ can be used to compute the averages of observables (see e.g.~\cite{Salvio:2025rma} for explicit formul\ae~for general equilibrium density matrices).

   Like in scalar-fermion theories previously discussed in~\cite{Salvio:2025rma,Salvio:2025ggj}, to account for a non-vanishing average angular momentum one has to substitute  
  $H_c$ with $H_c -\vec \Omega\cdot \vec J_c$ in the path integral, which is nothing but the transformation rule of the classical Hamiltonian from an inertial frame to a frame rotating with the angular-velocity vector $\vec \Omega$.      One can, thus, identify $\vec \Omega$ with the angular-velocity vector of the rotating plasma. One should keep in mind, however, that in TFT $\vec \Omega$ is a thermodynamical quantity, which allows us to compute the average value of the angular momentum  in any rotating plasma at thermal equilibrium.

To include matter fields, one can now combine in a simple way~(\ref{PIgauge}) and~(\ref{GenPartGauge}) with the results in~\cite{Salvio:2025rma,Salvio:2025ggj} to obtain a path integral expression for $Z$ and Green's functions of  the form in~(\ref{GreenO}), but where the operators $\mathcal{O}_i$ can now involve both gauge fields and matter fields.
If matter fields are also present one should also integrate over them with the appropriate measures, add to $\dot A_i(x)\Pi^i(x)$ the sum of the products of the time derivative of the canonical matter ``coordinates" times the conjugate matter ``momenta" and 
 add to  $\mathcal{H}_c^\omega$ the matter $\Omega$-dependent Hamiltonian density. Both the matter  measures and the matter $\Omega$-dependent Hamiltonian density were provided  in~\cite{Salvio:2025rma,Salvio:2025ggj}.

Let us see now how to handle the secondary constraint on $A_0^a$ with the path-integral approach.
Since $A_0^a$ appears at most quadratically in $\mathcal{H}_c^\omega$  and the quadratic piece $-\frac12 \partial_3 A^a_0\partial_3 A^a_0$ has a field-independent coefficient (see Eq.~(\ref{HcOmegagauge})), one can effectively treat  $A_0^a$ as an independent variable and path integrate over it:
\bea \langle {\cal T}\mathcal{O}_1(x_1) ...  \mathcal{O}_n(x_n) \rangle = \frac1{``O_i\to 1"}&&\hspace{-0.6cm}\int \delta A_0 \delta A_1\delta A_2 \, \delta \Pi^1\delta \Pi^2  \, O^\omega_1(x_1^\omega) ... O_n^\omega(x_n^\omega)\nonumber  \\ &&\hspace{-0.6cm}\exp\left(i \int_C d^4x \left(\dot A_i(x)\Pi^i(x) - \mathcal{H}^\omega_c(A(x), \Pi(x))\right)\right),\label{PIgauge2}\hspace{-0.3cm}\eea
where 
 \be \delta A_0 = \prod_{x, a} dA^a_0(x).\ee
Indeed, path integrating over $A_0^a$ such Gaussian function of  $A_0^a$  sets $A_0^a$ equal to the solution of its  field equations and produces a field-independent coefficient that cancels with the denominator $``O_i\to 1"$ (which coincides with the numerator for $O^\omega_1(x_1^\omega) ... O_n^\omega(x_n^\omega)\to 1$). To see that this is true even if the $O_i$ depend on $A_0^a$, one can first introduce a generating functional  coupling all fields to external currents, including $A_0^a$. The result of the path integration over $A_0^a$ would then set $A_0^a$ equal to the solution of its  field equations including the external-current contribution. The Green's functions are then generated by taking functional derivatives with respect to the external currents and then setting those currents to zero, which sets $A_0^a$ equal to the solution of its  field equations without the external-current contribution.

With $A_0^a$ regarded as an independent integration variable, $\mathcal{H}_c^\omega$ is clearly at most quadratic in the $\Pi^i_a$ with the quadratic piece $\frac{1}{2}\Pi_a^{i} \Pi_a^{i}$ featuring a field-independent coefficient (see again Eq.~(\ref{HcOmegagauge})). The corresponding Gaussian path integral over $\Pi^i_a$ sets 
\be \Pi^i_a =  F^{a}_{0i} - \Omega (\partial_\phi A_i^a+\epsilon_{ij} A^{a}_j)\ee 
and produces a field-independent coefficient that cancels again with the denominator $``O_i\to 1"$.
This sets the integrand in the exponential in~(\ref{PIgauge2}) equal to the Lagrangian density 
$\mathscr{L}_\omega$, which is obtained by substituting  in  $-\frac14 F^a_{\mu\nu}F^{a\mu\nu}$ \be \partial_t A_k^a\to\partial_t A_k^a - \Omega (\partial_\phi A_k^a+\epsilon_{kl3} A^{a}_l) \label{subL}\ee
(for all $k,l=1,2,3$). 
The path integral in~(\ref{PIgauge2}) can then be expressed in the Lagrangian form:
\bea \langle {\cal T}\mathcal{O}_1(x_1) ...  \mathcal{O}_n(x_n) \rangle = \frac1{``O_i\to 1"}&&\hspace{-0.6cm}\int \delta A \, \delta(A_3)  \, O^\omega_1(x_1^\omega) ... O_n^\omega(x_n^\omega) \exp\left(i \int_C d^4x \mathscr{L}_\omega\right),\label{PIgauge3}\hspace{-0.3cm}\eea 
where the functional delta function
\be \delta(A_3) \equiv \prod_{x, a}\delta(A^a_3(x)) \ee
enforces the axial gauge and
 \be \delta A = \delta A_0\delta A_1 \delta A_2 \delta A_3, \qquad \delta A_3 = \prod_{x, a} dA^a_3(x).\ee

Note that, since going from a non-rotating to a rotating frame corresponds to substituting $H\to H-\vec\Omega \cdot \vec J$, the convergence of the partition function  requires $H-\vec\Omega \cdot \vec J$ (at least for $\mu_a=0$) and the real part of the Euclidean action to be bounded from below. 
Moreover, at least on a finite lattice, the convergence of the ensemble averages of all operators imply the convergence of all Euclidean thermal Green's functions.

Note now that the same action 
$\int_C d^4x\mathscr{L}_\omega$ appearing in~(\ref{PIgauge3}) can be equivalently obtained by starting from $-\frac14 \int_C d^4x F^a_{\mu\nu}F^{a\mu\nu}$ and changing the gauge four-vector as
\be A^a (t,\vec x) \to \mathscr{R}^{-1}(t\vec\Omega)A^a (t,R(t\vec\Omega)\vec x)\equiv \mathscr{R}^{-1}(t\vec\Omega)A^a (x^\omega), \ee 
where $\mathscr{R}(t\vec \Omega)$ is the $4\times 4$ matrix, which represents the rotation of an angle $t\Omega$ around $\vec\Omega$ on four-vectors.
 To show the equivalence note that the  change of the spatial integration variable in the action from $\vec x$ to $R^{-1}(t\vec \Omega)\vec x$ has unit Jacobian for any  $t$.
So one sees again that having a plasma with a non-vanishing thermal vorticity $\vec\tau \equiv -\beta \vec\Omega$ is equivalent to going from an inertial frame to a frame that is rotating with angular-velocity vector $\vec \Omega$. 
This also implies that $\mathscr{L}_\omega$ is gauge invariant.  Therefore, as long as the operators $\mathcal{O}_i$ are also gauge invariant, we can use the Faddeev-Popov argument to rewrite the thermal Green's function in a generic gauge:
 \be \hspace{-0.7cm}\boxed{\langle {\cal T}\mathcal{O}_1(x_1) ...  \mathcal{O}_n(x_n) \rangle = \frac1{``O_i\to 1"}\int \delta A \, D_f(A) \delta(f(A)) \, O^\omega_1(x_1^\omega) ... O_n^\omega(x_n^\omega) \exp\left(i \int_C d^4x \mathscr{L}_\omega\right),}\hspace{-0.6cm}\label{PIgauge4}\ee 
where the function $f$ identifies the generic gauge condition through $f_a(A,x)=0$ (for example, in the axial gauge $f_a(A,x) = A^a_3(x)$), 
\be  \delta(f(A)) \equiv \prod_{x, a}\delta(f_a(A,x)) \ee
and $D_f(A)$ is the determinant of the ``matrix"
\be M_{ax,by}(A) \equiv \left.\frac{\delta f_a(A^\alpha,x)}{\delta \alpha^b(y)}\right|_{\alpha =0}, \ee 
with $A^\alpha$ being the gauge-field configuration  that is obtained from the gauge-field configuration  $A$ acting with the element of the gauge group with parameters $\alpha^a$.
 As usual,  we can generically substitute $D_f(A)$ with an integral over Grassmann field variables, $c^a$ and $\bar c^a$ (Faddeev-Popov ghosts):
\be D_f(A) \to \int \delta\bar c\,  \delta c \, \exp\left(i\int_C d^4x d^4y \, \bar c^a(x) M_{ax,by}(A) c^b(y)\right),  \label{GhostInt}\ee
where 
\be \delta c = \prod_{x, a} dc^a(x), \qquad \delta \bar c = \prod_{x, a} d\bar c^a(x). \ee 
Since $M_{ax,by}(A)$ acts on a space of functions satisfying the twisted periodicity boundary conditions~(\ref{PerConA}), the Faddeev-Popov ghosts obey the same conditions although they are Grassmann variables. Faddeev-Popov ghosts will be interacting with some other fields any time $D_f(A)$ depends on those fields. This is not the case, for example, in the axial gauge, so  $D_f(A)$ cancels with the denominator $``O_i\to 1"$ in that gauge and~(\ref{PIgauge3}) is recovered. 

We can generate the thermal Green's function by performing functional derivatives of the generating functional 
 \be \boxed{ \mathcal{Z} (J) =  \frac1{``J\to 0"} \int \delta A \, D_f(A) \delta(f(A)) \, \exp\left(i \int_C d^4x \left( \mathscr{L}_\omega + J(x)\mathscr{R}^{-1}(t\vec\Omega) A(x^\omega)\right)\right),\label{PIgenGauge}} \ee
 where the denominator ``$J \to0$" is the numerator for $J \to0$.
 
Now, let us make a remark on the extension of~(\ref{PIgauge4}) to include matter fields. If matter fields are also present  one should also integrate over them with the appropriate measures and add to  $\mathscr{L}_\omega$ the matter $\Omega$-dependent Lagrangian density. The scalar and fermion measures, $\delta\varphi$ and $\delta\bar\psi\delta\psi$ respectively, and the matter $\Omega$-dependent Lagrangian density can be found in the previous works~\cite{Salvio:2025rma,Salvio:2025ggj}.

 The axial gauge was very useful to derive the path-integral representation of the thermal Green's functions from first principles. However, now one can choose an arbitrary gauge to perform calculations. 
 Another possible convenient choice for the gauge-fixing function  is \be f_a(A,x) = (\partial_0 - \Omega \partial_\phi )A_0^a(x)-\partial_kA_k^a(x) +\tau_g^a(x),\label{tauGauge}\ee
 where $\tau_g^a$ are fixed fields and different choices of them are thus different gauges.  However, $M_{ax,by}(A)$ and $D_f(A)$ do not depend on $\tau_g$. Explicitly, with the gauge choice in~(\ref{tauGauge}), the ghost insertion in the path integral reads
\be \hspace{-0.4cm}D_f(A) \to \int \delta\bar c\,  \delta c \, \exp\left(i\int_C d^4x \, \bar c^a(x) \left\{\delta_{ab}\Box_\omega-gf^{abc}\left[(\partial_0 - \Omega \partial_\phi )A_0^c(x)-\partial_kA_k^c(x)\right]\right\} c^b(x)\right),  \label{GhostInttau}\ee
with
\be \Box_\omega  = \left(\partial_t - \Omega \partial_\phi\right)^2-\partial_k\partial_k,  \ee
the same differential operator appearing in the quadratic scalar-field action (see Ref.~\cite{Salvio:2025rma}). Both the quadratic part and the interaction terms in the Faddeev-Popov ghost Lagrangian depend on $\Omega$ with the gauge choice in~(\ref{tauGauge}). In Sec.~\ref{Interactions}, however, we will see that in perturbation theory only the propagators are $\Omega$-dependent, the vertices are not and are equal to those with $\Omega=0$ and $\mu_a=0$.

One can exploit the fact that both the numerator and the denominator in~(\ref{PIgauge4}) are gauge invariant and thus independent of $\tau_g$ and multiply both of them by the path integral over $\tau_g^a$ of the exponential $ \exp(-i\int_C d^4x \, \tau_g^a(x)\tau_g^a(x)/(2\xi_g))$. Using then $\delta(f(A))$ to perform the path integral over $\tau_g$ the Lagrangian density of the gauge fields is changed as follows
\be\mathscr{L}_\omega\to \mathscr{L}_\omega -\frac{1}{2\xi_g} \sum_a\left[(\partial_t - \Omega \partial_\phi )A_0^a(x)-\partial_kA_k^a(x)\right]^2. \label{xiLag} \ee 
For $\xi_g\to0$ one obtains what can be called the rotating Landau gauge, $(\partial_t - \Omega \partial_\phi )A_0^a(x)-\partial_kA_k^a(x)=0$,
while one could call the $\xi_g=1$ choice the rotating Feynman gauge, which leads to some simplification in perturbation theory, as shown in Sec.~\ref{propGauge}.

\section{Generalized Kubo-Martin-Schwinger conditions}\label{Generalized Kubo-Martin-Schwinger condition}

The Kubo-Martin-Schwinger (KMS) condition is an exact (non-perturbative) relation between the two non-time-ordered 2-point functions of a pair of field operators, $\mathcal{O}_1$ and $\mathcal{O}_2$, in a finite-temperature field theory~\cite{KMS}. Here this condition is extended to include generic values of both the $\mu_a$ and $\Omega$, keeping the  Lorentz and internal-symmetry representations for $\mathcal{O}_1$ and $\mathcal{O}_2$ general (generalized KMS condition). So one can apply these conditions to {\it any} TFT with equilibrium density matrix. Both the coordinate- and momentum-space generalized KMS conditions will be derived. 

\subsection{Coordinate space}
Exploiting the cyclicity of the trace one finds
\bea &&G^<(x_1,x_2)\equiv \langle \mathcal{O}_2(x_2)\mathcal{O}_1(x_1)\rangle \equiv \frac1{Z} \Tr\left(e^{-\beta (H-\vec\Omega  \cdot \vec J - \mu_a Q^a)}\mathcal{O}_2(x_2)\mathcal{O}_1(x_1)\right)\nonumber \\
 &=&  \frac1{Z} \Tr\left(e^{-\beta (H-\vec\Omega  \cdot \vec J - \mu_a Q^a)} e^{\beta (H-\vec\Omega  \cdot \vec J - \mu_a Q^a)}\mathcal{O}_1(x_1)e^{-\beta (H-\vec\Omega  \cdot \vec J - \mu_a Q^a)} \mathcal{O}_2(x_2)\right).\eea 
On the other hand, one can write 
\be e^{\beta (H-\vec\Omega  \cdot \vec J - \mu_a Q^a)}\mathcal{O}_1(x_1)e^{-\beta (H-\vec\Omega  \cdot \vec J - \mu_a Q^a)}  = e^{-\beta \mu_a T^a_1} \mathcal{D}_1(-i\beta\vec\Omega)\mathcal{O}_1(t_1-i\beta,R(i\beta \vec \Omega)\vec x_1),\ee 
where the $T_1^a$ are the generators of the internal-symmetry group in the representation of $\mathcal{O}_1$ (note then that the $T_1^a$ commute with $\mathcal{D}_1$, which acts instead on Lorentz indices),
to obtain 
\be \boxed{G^<(x_1,x_2) = e^{-\beta \mu_a T^a_1} \mathcal{D}_1(-i\beta\vec\Omega)G^>(t_1-i\beta,R(i\beta\vec \Omega)\vec x_1,x_2)
,} \label{KMSc}\ee
where $G^>(x_1,x_2)\equiv G^>(t_1,\vec x_1, x_2)\equiv \langle \mathcal{O}_1(x_1)\mathcal{O}_2(x_2)\rangle$.
Eq.~(\ref{KMSc}) represents the desired generalized KMS conditions in coordinate space.

Note the generality of the result: not only the values of the $\mu_a$ and $\Omega$ but also the Lorentz and internal-symmetry representations  are arbitrary. For example, if $\mathcal{O}_1$ is a scalar operator $\mathcal{D}_1=1$, while, for four-vector operators $\mathcal{D}_1(t\vec\Omega)=\mathscr{R}(t\vec \Omega)$.
To the best of our knowledge, such a general form of the KMS conditions has never been determined before (see, however, Refs.~\cite{Landsman:1986uw,Quiros:1994dr,Bellac:2011kqa,Ambrus:2021eod} for specific cases).

\subsection{Analyticity}\label{Analyticity}

The generalized KMS conditions in~(\ref{KMSc}) involve  a non-time-ordered 2-point function evaluated at complex values of its arguments, $G^>(t_1-i\beta,R(i\beta\vec \Omega)\vec x_1,x_2)$. One can show that if $G^>(x_1,x_2)$ is a well-defined function  for real $x_1$ and $x_2$ and $\beta\geq0$ then it remains well defined after the substitution $\{t_1,\vec x_1\}\to \{t_1-i\beta,R(i\beta\vec \Omega)\vec x_1\}$.

 To prove this analyticity property let us use the cylindrical coordinates in~(\ref{CyCoo}) to write $\{x^1_1, x^2_1, x^3_1\}=\{r_1\cos\phi_1, r_1\sin\phi_1, z_1\}$ and $\{x^1_2, x^2_2, x^3_2\}=\{r_2\cos\phi_2, r_2\sin\phi_2, z_2\}$ and neglect irrelevant rotation matrices around the third axis of angles $\phi_1$ and $\phi_2$. One can then express the 2-point function in question in terms of the following spectral decomposition
 \bea \Tr\left(\rho\,e^{i(Ht-\phi J_3)}\mathcal{O}_1(0,r_1,0,z)e^{-i(Ht-\phi J_3)}\mathcal{O}_2(0,r_2,0,0)\right)\nonumber\\
 \hspace{-0cm}=\frac{1}{Z} \sum_{qq'}\langle q'|e^{\beta \mu_a Q^a}\mathcal{O}_1(0,r_1,0,z)|q\rangle\langle q|\mathcal{O}_2(0,r_2,0,0)|q'\rangle e^{i\omega_{q'}(t+i\beta)-im_{q'}(\phi+i\beta\Omega)}e^{-i \omega_q t+im_q \phi}, \label{AnalyStep}\eea
with $t\equiv t_1-t_2$, $\phi\equiv \phi_1-\phi_2$, $z\equiv z_1-z_2$ and the states $|q\rangle$ form a complete set of eigenstates of $H$ and $J_3$, namely $H|q\rangle = \omega_q|q\rangle$ and $J_3|q\rangle = m_q|q\rangle$. As stated above, one is assuming here that this series converges for $t$ and $\phi$ real and $\beta\geq 0$.  Then it continues to converge even after substituting
\be t\to t-i\tau, \qquad \phi\to \phi-i\tau \Omega, \label{subtphi}\ee
with $\tau \in [0,\beta]$. To see this one can use the equation (which follows from the transformation rule of $\mathcal{O}_1$ under $\mathcal{G}$)
\be \mathcal{O}_1 = e^{\tau \mu_a T^a_1}e^{-\tau \mu_a Q^a} \mathcal{O}_1e^{\tau \mu_a Q^a}\ee
at the same time as the substitution above  and take $|q\rangle$ to be an eigenstate also of $\mu_aQ^a$ with eigenvalue $\mathcal{M}_q$. In this way one obtains
\bea \frac{e^{\tau \mu_a T^a_1}}{Z} \sum_{qq'}\langle q'|e^{(\beta-\tau) \mu_a Q^a}\mathcal{O}_1(0,r_1,0,z)|q\rangle\langle q|\mathcal{O}_2(0,r_2,0,0)|q'\rangle \nonumber \\ \times \,e^{i\omega_{q'}(t+i(\beta-\tau))-im_{q'}(\phi+i(\beta-\tau)\Omega)}e^{-i \omega_q t+im_q \phi}e^{-\tau(\omega_q-\Omega m_q -\mathcal{M}_q)}\eea 
and, modulo a harmless overall matrix, $\exp(\tau \mu_a T^a_1)$, one reproduces the initial form of the terms of the series with the substitution $\beta\to \beta -\tau$ and the extra exponential factor $\exp(-\tau(\omega_q-\Omega m_q-\mathcal{M}_q))$. For $\tau>0$ this factor generically helps the convergence of the series in order to ensure that the ensemble averages and $Z$  are finite. Since $\tau \in [0,\beta]$, the  analyticity property mentioned above is thus proved.

\subsection{Momentum space}

Armed with the result of Sec.~\ref{Analyticity}, we can derive  the generalized KMS conditions for the (partial) Fourier transforms with respect to the coordinates $\xi\equiv \{t, z, \phi\}$:
\be  \tilde G^>(k; r_1, r_2) \equiv \int d^3 \xi \, e^{ik\xi}  \, G^>(\xi; r_1, r_2), \qquad \tilde G^<(k; r_1, r_2) \equiv \int d^3 \xi \, e^{ik\xi}  \, G^<(\xi; r_1, r_2).\ee
Here we are using the fact that $G^>(x_1, x_2)$ and $G^<(x_1, x_2)$ are in fact functions only of $t\equiv t_1-t_2$, $\phi\equiv \phi_1-\phi_2$, $z\equiv z_1-z_2$ and $r_1$ and $r_2$ and so they can be expressed in the form $G^>(\xi; r_1, r_2)$ and $G^<(\xi; r_1, r_2)$, respectively. Also, the notation $k\xi \equiv k_0 t - p z -m \phi$ is used.

Applying the generalized coordinate-space KMS conditions in~(\ref{KMSc}), one finds
\bea  &&\tilde G^<(k; r_1, r_2) = \int d^3\xi \, e^{ik\xi} \, e^{-\beta \mu_a T^a_1} \mathcal{D}_1(-i\beta\vec\Omega)G^>(t_1-i\beta, R(i\beta\vec \Omega)\vec x_1,x_2) \nonumber \\
 &=&e^{-\beta (k_0-m\Omega)} e^{-\beta \mu_a T^a_1} \mathcal{D}_1(-i\beta\vec\Omega) \int dz\int_{-\infty-i\beta}^{+\infty-i\beta} dt \int_{-i\beta \Omega}^{2\pi-i\beta \Omega} d\phi \, e^{ik\xi} \,G^>(\xi; r_1, r_2). \eea
The analyticity established in Sec.~\ref{Analyticity} then tells us
 \be \boxed{\tilde G^<(k; r_1, r_2) = e^{-\beta (k_0-m\Omega)} e^{-\beta \mu_a T^a_1} \mathcal{D}_1(-i\beta\vec\Omega) \tilde G^>(k; r_1, r_2),}  \ee
which represents  the generalized KMS conditions in momentum space. This version  of the generalized KMS conditions can be used, among other things, to easily relate absorption and production rates of particles coupled to the rotating plasma.

\section{Thermal propagators}\label{propGauge}

Another application of the generalized KMS conditions is  the computation of the thermal propagators. These are essential ingredients in perturbation theory and we thus compute them here for a general gauge theory in the presence of arbitrary values of the $\mu_a$ and $\Omega$. 


\subsection{Warming up with scalars}\label{Warming up with scalars}

As a first step let us consider a set of scalars $\varphi_s$ in an arbitrary irreducible representation of the internal symmetry group and with generating functional~\cite{Salvio:2025rma} \be \mathcal{Z}(j) = \frac1{``j\to 0"} \int \delta\varphi \exp\left(iS_C(\varphi)+i\int_C d^4x j(x)\varphi(x^\omega)\right),  \label{GenFunLag} \ee
   where the denominator ``$j \to0$" is the numerator for vanishing scalar external current, $j(x) \to0$, and
   \be  S_C(\varphi) = \int_C d^4x \left(-\frac12 \varphi (\Box_\omega+\mu^2)\varphi - \mathcal{V}(\varphi)\right). \label{SC}\ee
Here $\mu$ is the mass of the scalars in the irreducible representation in question. 
 
  Indeed, much of the functional techniques that will be developed here for scalars will be used later on to determine the thermal propagators in the gauge sector (including both gauge fields and Faddeev-Popov ghosts).

The thermal propagators can be computed with the stationary-point method and the use of the generalized KMS conditions, used as boundary conditions, as shown in this section.

By performing a rotation of angle $t\Omega$ around $\vec\Omega$ of the spatial integration variable (which has unit Jacobian)
  to rewrite the coupling with the external current in the action as 
\be \int_C d^4x j(x)\varphi(x^\omega) = \int_C d^4x j(x^{-\omega})\varphi(x),  \label{ChangSpI}\ee
where $x^{-\omega}\equiv \{t, R(-t\vec\Omega)\vec x\}$, one finds that the stationary point $\varphi_0$ of the argument of the exponential in the path integral satisfies 
\be (\Box_\omega+\mu^2)\varphi_0 = j(x^{-\omega}). \label{stationaryScalar}\ee
The solution of this equation is
\be \varphi_0(x) = -\int_C d^4y D(x,y)  j(y^{-\omega}),\ee
where $D$ here is the Green's function of  $-(\Box_\omega+\mu^2)$:
\be -(\Box_\omega^x+\mu^2) D(x,y)  = \delta(x-y) \label{EqGFD} \ee
(the differential operator $\Box_\omega^x$ is $\Box_\omega$ acting on the spacetime variable $x$). Since $(\Box_\omega^x+\mu^2)\delta(x-y)$ is symmetric, meaning that exchanging $x\leftrightarrow y$ does not change it, one finds the symmetry property $D_{ss'}(x,y)=D_{s's}(y,x)$. 

One can then easily show with functional methods that the (free) thermal propagator is 
\be \langle {\cal T}\Phi_s(x_1)\Phi_{s'}(x_2)\rangle=iD_{ss'}(x_1^\omega,x_2^\omega), \ee
where the $\Phi_s$ are the quantum fields corresponding to $\varphi_s$.
The fact that $D$ here is computed in $\{x_1^\omega,x_2^\omega\}$ rather than in $\{x_1,x_2\}$ is because the external current  in~(\ref{ChangSpI}) and~(\ref{stationaryScalar}) is computed in $x^{-\omega}$  rather than in $x$. This leads to 
\be -(\Box^x+\mu^2) D(x^\omega,y^\omega)  = \delta(x-y),\label{EqGFDo} \ee
with $\Box^x$ being the d'Alembertian operator acting on the spacetime variable $x$.
So the equation that the thermal propagator satisfies does not depend on the $\mu_a$ and $\Omega$. The dependence on these quantities will be introduced by the generalized KMS conditions.  

Eq.~(\ref{EqGFD}) suggests that $D(x_1,x_2)$ can in fact, just like $G^>(x_1, x_2)$ and $G^<(x_1, x_2)$, be expressed in cylindrical coordinates in terms only of $t\equiv t_1-t_2$, $\phi\equiv \phi_1-\phi_2$, $z\equiv z_1-z_2$ and $r_1$ and $r_2$; this is indeed true because it is implied by the fact that $iD(x_1^\omega,x_2^\omega)$ should be the thermal  propagator. Then 
\be \langle {\cal T}\Phi_s(x_1)\Phi_{s'}(x_2)\rangle=iD_{ss'}(\xi^\omega; r_1, r_2), \ee
where $\xi^\omega\equiv \{t, z, \phi-\Omega t\}$ and Eq.~(\ref{EqGFDo}) becomes
\be \left(-\partial_t^2+\partial_{r_1}^2+\frac1{r_1}\partial_{r_1}+\frac1{r_1^2}\partial_{\phi}^2+\partial_z^2-\mu^2\right)D(\xi^\omega; r_1, r_2) = \delta(t)\delta(\phi)\delta(z)\frac{\delta(r_1-r_2)}{r_1}. \ee
We can then write   
\be D(\xi^\omega; r_1, r_2) = \sum_{m=-\infty}^{+\infty}\frac{e^{im\phi}}{2\pi}\int_{-\infty}^{\infty}\frac{dp}{2\pi}e^{ipz}\int_0^\infty d\alpha\alpha  J_m(\alpha r_1)J_m(\alpha r_2) D_s(t,p,m,\alpha), \label{Dxio}\ee
where $D_s$ is some function of $t, p, m$ and $\alpha$.

When the system is enclosed in a cylinder of radius $R$ we should also impose the continuity condition $J_m(\alpha R)=0$ on the $J_m$, which implies that $\alpha$  can only assume the discrete values 
  \be \alpha_{m,n} \equiv \frac{j_{m,n}}{R},\qquad n = 1,2,3,...\,,  \label{alphadef}\ee
   where $j_{m,n}$ is the $n$th positive zero of the function $J_m$. Since a finite $\Omega$ implies a finite $R$, the variable $\alpha$ only assumes discrete values and one should actually read an integral as a sum over discrete values:
   \be \int_0^\infty d\alpha\alpha \left[...\right] =\sum_{n=1}^\infty  \Delta \alpha_{m,n} \, \alpha_{m,n} \left[...\right]_{\alpha\to\alpha_{m,n}}, \label{DicMes} \ee
where  $\Delta \alpha_{m,n}\equiv \alpha_{m,n+1}-\alpha_{m,n}$, both in~\eqref{Dxio} and in the following equations. This ensures that~\eqref{vrange} is satisfied for a finite $\Omega$.

Nevertheless, the continuous-$\alpha$ approximation  can still be useful in physical situations in which rotational effects are present. Eq.~\eqref{alphadef} tells  us that this approximation corresponds to the large $R$ limit and therefore, according to~\eqref{vrange}, to the small $\Omega$ limit. This limit, however, can be taken with a fixed non-vanishing value of $v\in(0,1)$.  So in this approximation one trades the angular velocity $\Omega$ with the velocity $v$ at the radial boundary. In Refs.~\cite{Salvio:2025rma,Salvio:2025ggj} it was shown that a non-zero $v$ can lead to finite physical effects of rotation even if $\Omega$ is infinitesimal: for example, one can obtain a non-vanishing value of the average angular momentum along the $z$ axis~\cite{Salvio:2025rma,Salvio:2025ggj} and to finite rotational effects on particle-production rates
per unit of volume~\cite{Salvio:2025rma}. A physical interpretation of a continuous $\alpha$ will be provided below. 
   
Using now the completeness relation of the Bessel functions
\be \int_0^\infty d\alpha\alpha  J_m(\alpha r_1)J_m(\alpha r_2) = \frac{\delta(r_1-r_2)}{r_1}, \ee
and their closure equation one obtains 
\be \left(-\partial_t^2-\omega^2(\alpha,p)\right)D_s(t,p,m,\alpha) = \delta(t), \label{EqDs} \ee 
with $\omega(\alpha,p)\equiv \sqrt{\mu^2+\alpha^2+p^2}$.
On the other hand, from the definition of the time-ordered product ${\cal T}$ one can write
 \be D_s(t,p,m,\alpha) = D^>_s(t,p,m,\alpha)\theta(t)+D^<_s(t,p,m,\alpha)\theta(-t),\ee 
 where $\theta(x)$ is the Heaviside step function,
 to obtain (inserting in~(\ref{EqDs}))
 \be D_s^{>(<)}(t,p,m,\alpha) = D_p^{>(<)}(p,m,\alpha)e^{-i\omega(\alpha,p) t}+D_n^{>(<)}(p,m,\alpha)e^{i\omega(\alpha,p) t}, \ee
 \be D_p^>=D_p^<-\frac{i}{2\omega}, \qquad D_n^>=D_n^<+\frac{i}{2\omega},\label{PropStep2}\ee 
 where $D_p^{>(<)}$ and $D_n^{>(<)}$ are some functions of $p, m$ and $\alpha$.
 
 The coefficients $D_p^{>(<)}$ and $D_n^{>(<)}$ can now be determined using the generalized KMS conditions in~(\ref{KMSc}). These conditions imply, recalling that we are dealing with scalars,
 \be D_p^< = e^{-\beta (\omega-m\Omega+\mu_aR^a)}D_p^>, \qquad D_n^< = e^{\beta (\omega+m\Omega-\mu_aR^a)}D_n^>, \label{PropStep3} \ee 
 with $R^a$ being the generators of $\mathcal{G}$ in the given irreducible representation. Combining~(\ref{PropStep2}) and~(\ref{PropStep3}) then leads to the solutions
 \be D_p^> = -\frac{i}{2\omega} (1+f_B (\omega-m\Omega+\mu_aR^a)), \qquad D_n^> = -\frac{i}{2\omega} f_B (\omega+m\Omega-\mu_aR^a),\ee 
 where
\be f_B(x) \equiv \frac1{e^{\beta x}-1} \label{fBdef}\ee
is the Bose-Einstein distribution.

 Finally, the expression for the free  non-time-ordered 2-point function is given by 
 \bea G^>(x_1, x_2) &=& \sum_{m=-\infty}^{+\infty}\int_\mu^\infty d\omega\int_{-p_0}^{p_0}dp \left[ (1+f_B (\omega-m\Omega+\mu_aR^a))e^{-i\omega t+ipz+im\phi} \frac{J_m(\alpha r_1)J_m(\alpha r_2)}{2^3\pi^2}\right. \nonumber \\
&&\left.+ f_B (\omega-m\Omega-\mu_aR^a)e^{i\omega t-ipz-im\phi} \frac{J_m(\alpha r_1)J_m(\alpha r_2)}{2^3\pi^2}\right],\label{GforS}\eea 
where $\alpha \equiv \sqrt{p_0^2-p^2}$, 
$p_0\equiv \sqrt{\omega^2 -\mu^2}$. 
The other non-time-ordered 2-point function  can be obtained in a straightforward way by using the property $G_{ss'}^<(x_1,x_2) = G_{s's}^>(x_2,x_1)$. Of course, one can then easily combine $G^>$ and $G^<$ to obtain the thermal propagator: $$\langle {\cal T}\Phi_s(x_1)\Phi_{s'}(x_2)\rangle=\theta(t_1-t_2)G_{ss'}^>(x_1,x_2)+\theta(t_2-t_1)G_{ss'}^<(x_1,x_2). \label{PropScalar}$$ 

Enclosing  the system in a cylinder of radius $R$,  the discrete values of $\alpha$ in~\eqref{alphadef} correspond to the discrete values of $\omega$ given by
\be \omega_{m,n}(p) \equiv \sqrt{\mu^2+ \alpha_{m,n}^2 + p^2},\qquad n = 1,2,3,...\,  \label{omegamn} \ee
and in~\eqref{GforS} one should read an integral as a sum over discrete values:
\be \sum_{m=-\infty}^{+\infty}\int_\mu^\infty d\omega\int_{-p_0}^{p_0}dp [...] = \sum_{m=-\infty}^{+\infty} \int_{-\infty}^{\infty}dp \sum_{n=1}^\infty  \Delta \omega_{m,n}(p) \left[...\right]_{\omega\to\omega_{m,n}(p)},  \ee
where 
\be \Delta \omega_{m,n}(p) \equiv \frac{\alpha_{m,n}}{\omega_{m,n}(p)}\Delta\alpha_{m,n}. \ee
This interpretation ensures that~\eqref{vrange} is satisfied for a finite $\Omega$. However, as already mentioned, the approximation in which $\alpha$ (and thus $\omega$)  is really treated as a continuous variable can still be useful in physical situations in which rotational effects are present. Let us give a physical interpretation of this approximation. In the exponent of the density matrix $\omega$  appears multiplied by $1/T$. So substituting $\omega$ with a continuous variable is valid when $T \gg 1/R$. This condition is typically satisfied by macroscopic objects at room temperature and amply satisfied by astrophysical objects (interpreting $R$ as the linear size of the object). Since $1/R>\Omega$ (Eq.~\eqref{vrange}), the condition $T \gg 1/R$ implies $T \gg \Omega$. As a result, the approximation in which $\alpha$ (and thus $\omega$)  is really treated as a continuous variable corresponds to a small value of $\Omega/T$. 

 The expression in~\eqref{GforS} reproduces the propagator found in~\cite{Salvio:2025rma}, where the times  $t_1$ and $t_2$ were taken on the real axis and creation and annihilation operators were used. However, the formula found here holds not only when $t_1$ and $t_2$ are taken on the real axis, but  for any $t_1$ and $t_2$ on $C$. Moreover, the use of functional techniques will be essential in this paper to study the full gauge sector, including gauge fields and Faddeev-Popov ghosts, for the reasons explained at the end of Sec.~\ref{Preliminary remarks on free spin-1 particles}.
 
 The quantity $f_B(\omega - m\Omega \pm \mu_a R^a)$ appearing in~(\ref{GforS}) is the Bose-Einstein distribution $f_B$ in~(\ref{fBdef}) computed at the matrix $\omega-m\Omega\pm \mu_a R^a$ (where an identity matrix multiplying $\omega-m\Omega$ is understood). In practice, to compute this quantity one can determine the projectors $\mathcal{P}_d$ associated with the eigenvalues $\mathcal{M}_d$ of  $\mu_a R^a$ and use the well-known spectral decomposition
\be f_B(\omega - m\Omega \pm \mu_a R^a) = \sum_d f_B(\omega - m\Omega \pm \mathcal{M}_d)\mathcal{P}_d, \label{SpecDec}\ee 
where the projectors always satisfy $\sum_d\mathcal{P}_d =1$, $\mathcal{P}_d\mathcal{P}_{d'} = \delta_{dd'}\mathcal{P}_{d'}$ and $\mu_a R^a = \sum_d\mathcal{M}_d \mathcal{P}_d$.  Both the eigenvalues $\mathcal{M}_d$ and the projectors $\mathcal{P}_d$ generically depend on the chemical potentials $\mu_a$ and are computable once the internal-symmetry group $\mathcal{G}$ and the irreducible representation with generators $R^a$ are specified.

\subsection{Gauge fields}

As explained at the end of Sec.~\ref{Quantization and general path-integral formula}, the Lagrangian density changes by changing the gauge in a generically $\Omega$-dependent way (see, for example,~Eq.~(\ref{xiLag})). This implies not only that the thermal propagators of the gauge fields can depend on the gauge, but also that some extra gauge-dependent $\Omega$ dependence can appear. The possible ways of fixing the gauge are infinite, so here we focus on a particularly convenient choice, the rotating Feynman gauge ($\xi_g=1$ in Eq.~(\ref{xiLag})). As we shall see here, this setting indeed leads to a simple form of the gauge-field thermal propagators that can in turn simplify the perturbative expansion.

In the rotating  Feynman gauge the quadratic Lagrangian density for gauge fields turns out to be
\be \frac12 A_0^a\Box_\omega A_0^a-\frac12 A_3^a\Box_\omega A_3^a-\frac12 A_i^a\Box_\omega A_i^a,
 \ee
which is obtained by considering only the terms quadratic in the gauge fields in the right-hand side of~(\ref{xiLag}) for $\xi_g=1$. The differential operator $\Box_\omega$ acts on $A_0^a$ and $A_3^a$ just like on scalar fields: it can be obtained from the d'Alembertian operator $\Box$ through the substitution $\partial_t\to \partial_t-\Omega\partial_\phi$. However,  $\Box_\omega$ acts on $A_i^a$ differently: its action on $A_i^a$ can be obtained from  $\Box A_i^a$ through the more involved substitution $\partial_tA_i^a\to \partial_tA_i^a-\Omega\partial_\phi A_i^a -\Omega \epsilon_{ij}A_j^a$. So the quadratic Lagrangian of $A_3^a$ (and $A_0^a$) is equal (respectively opposite) to that of massless scalars, while that of $A_i^a$ contains extra pieces due to its non-invariance under rotations around $\vec\Omega$.

Looking at~(\ref{PIgenGauge}), the coupling of $A_0$ and $A_3$ to the external current is the same as that of scalars. As a result, the thermal propagator of $A_3$ (and $A_0$) in this gauge coincides with (is respectively opposite to) that of a massless scalar in the adjoint representation of the gauge group, which is a particular case of the setup already analyzed in Sec.~\ref{Warming up with scalars}. 

 On the other hand,  the coupling of $A_i$ to the external current in the action can be rewritten as follows:
\be \int_Cd^4xJ^i(x)R_{ij}(-t\vec\Omega) A_j(x^\omega)=\int_Cd^4xR_{ji}(t\vec\Omega)J^i(x^{-\omega}) A_j(x), \ee
where a rotation of angle $t\Omega$ around $\vec\Omega$ of the spatial integration variables was performed. Note that this coupling, unlike the analogous coupling for scalars in~(\ref{ChangSpI}), includes the rotation matrix $R(t\vec\Omega)$. Adapting the procedure presented for scalar fields in Sec.~\ref{Warming up with scalars}, one finds that the presence of the rotation matrix in the coupling to the external current precisely cancels the presence of $\Omega$ in the equation that the thermal propagator satisfies, which then reads 
\be i\Box^x \langle {\cal T}A_i^a(x)A_j^b(y)\rangle  = \delta_{ij}\delta_{ab}\delta(x-y).\label{EqGFDo2} \ee 
Just like for scalars, that equation does not depend on the $\mu_a$ and $\Omega$. The dependence on these quantities will be introduced again by the generalized KMS conditions in~(\ref{KMSc}), which are, however, different for vector operators because of the presence of the matrix $\mathcal{D}_1(-i\beta\vec\Omega)$. For the $A_i$ this matrix is
\be \mathcal{D}_1(-i\beta\vec\Omega) = e^{\beta \Omega\sigma_2 }, \ee
where $\sigma_2$ is the second Pauli matrix. The KMS conditions can then be imposed in a way similar to that presented in Sec.~\ref{Warming up with scalars} for scalars, except that one should take into account this non-trivial matrix. The result for $G_{ij}^{>ab}(x_1,x_2)\equiv  \langle A_i^a(x_1)A_j^b(x_2)\rangle$ is
\bea G^>(x_1, x_2) &=& \sum_{m=-\infty}^{+\infty}\int_0^\infty d\omega\int_{-\omega}^{\omega}dp \left[ (1+f_B (\omega-(m+\sigma_2)\Omega+\mu_af^a))e^{-i\omega t+ipz+im\phi} \frac{J_m(\alpha r_1)J_m(\alpha r_2)}{2^3\pi^2}\right. \nonumber \\
&&\left.+ f_B (\omega-(m-\sigma_2)\Omega-\mu_af^a)e^{i\omega t-ipz-im\phi} \frac{J_m(\alpha r_1)J_m(\alpha r_2)}{2^3\pi^2}\right],\label{GFprop}\eea
with $f^a$ being the generators of the gauge group  in the adjoint representation. 

The other non-time-ordered 2-point function  can be easily obtained by using the property $G_{ij}^{<ab}(x_1,x_2) = G_{ji}^{>ba}(x_2,x_1)$ and one can then  obtain the thermal propagator: $$\langle {\cal T}A_i^a(x_1)A_j^b(x_2)\rangle=\theta(t_1-t_2)G_{ij}^{>ab}(x_1,x_2)+\theta(t_2-t_1)G_{ij}^{<ab}(x_1,x_2).$$ 

When the system is enclosed in a cylinder of radius $R$ only discrete values of $\alpha$ are allowed by the continuity condition $J_m(\alpha R)=0$ on the $J_m$.

Now, note that one can diagonalize $\sigma_2$ in~(\ref{GFprop}) by going  to the spin basis $A_\pm^a\equiv (A_1^a\pm iA_2^a)/\sqrt{2}$, where $ \langle A_\pm^a(x_1)A_\pm^b(x_2)\rangle = 0$ and $ \langle A_\pm^a(x_1)A_\mp^b(x_2)\rangle$ reads as~(\ref{GFprop}), except that $\sigma_2$ is substituted by $\mp 1$, respectively. Once the gauge group is specified, one can then compute $f_B$ at the matrices $\omega-(m\mp 1)\Omega+\mu_af^a$ and $\omega-(m\pm 1)\Omega-\mu_af^a$ by determining  the eigenvalues and the projectors of $\mu_af^a$  in a way analogous to that adopted for scalars around Eq.~(\ref{SpecDec}).  For example, if the gauge group is SU(3) with generators in the fundamental representation given by half the Gell-Mann matrices $\lambda^a$ (with $a=1,...8$) and one turns on only the color chemical potentials $\mu_3$ and $\mu_8$ (as suggested by studies of the superconducting color-flavor-locked phase of QCD~\cite{Alford:2002kj,Steiner:2002gx}), then $\mu_af^a$ has a  block-diagonal form
\be \mu_af^a = {\rm diag}\left(\mathcal{M}_+\sigma_2,0,\mathcal{M}^{(+)}_+\sigma_2,\mathcal{M}^{(-)}_- \sigma_2,0\right), \ee
with eigenvalues $\mathcal{M}_0=0$, $\mathcal{M}_{\pm} = \pm \mu_3$, $\mathcal{M}^{(+)}_{\pm} =  (\mu_3\pm\sqrt{3}\mu_8)/2$ and $\mathcal{M}^{(-)}_{\pm} =  -(\mu_3\pm\sqrt{3}\mu_8)/2$ and block-diagonal projectors respectively given by
\bea \mathcal{P}_0= {\rm diag}\left(0,0,1,0,0,0,0,1\right), \quad  \mathcal{P}_\pm= {\rm diag}\left((1\pm\sigma_2)/2,0,0,0,0,0,0\right), \nonumber \\
\mathcal{P}^{(+)}_+= {\rm diag}\left(0,0,0,(1+\sigma_2)/2,0,0,0\right), \quad \mathcal{P}^{(+)}_-= {\rm diag}\left(0,0,0,0,0,(1-\sigma_2)/2,0\right), \nonumber \\
\mathcal{P}^{(-)}_+= {\rm diag}\left(0,0,0,(1-\sigma_2)/2,0,0,0\right), \quad \mathcal{P}^{(-)}_-= {\rm diag}\left(0,0,0,0,0,(1+\sigma_2)/2,0\right), \eea
where $1$ in $1\pm\sigma_2$  is the $2\times2$ identity matrix.

To the best of our knowledge an explicit  closed-form expression for the thermal gauge-field propagator was never obtained before in the presence of $\Omega$. Note that here not only $\Omega$, but also an arbitrary number of chemical potentials is included.

\subsection{Faddeev-Popov Ghosts}

The thermal propagator of the Faddeev-Popov ghosts is gauge dependent, like that of gauge fields. Let us focus here on  the gauges~(\ref{tauGauge}), which include the $\xi_g$ gauges (and thus the Feynman gauge) as particular cases.

In those gauges, the differential operator appearing in the piece of the Lagrangian that is quadratic in the Faddeev-Popov Ghosts is the opposite of $\Box_\omega$ for scalars (see~(\ref{GhostInttau})). Furthermore,  the Faddeev-Popov ghosts  obey  the same twisted periodicity boundary conditions of gauge fields. As a result, the ghost thermal propagator in the gauges~(\ref{tauGauge}) is minus  the thermal propagator  of a massless scalar  in the adjoint representation of the gauge group. The same is true for the two non-time-ordered 2-point functions.

\section{Interactions}\label{Interactions}

The substitution in~(\ref{subL}) shows that  in a non-Abelian Yang-Mills theory $\Omega$
  affects not only the part of the action quadratic in the field variables, but also introduces the following additional cubic interaction in the Lagrangian density:
 \be \Omega gf^{bca}A^b_0 A^c_k(\partial_\phi A_k^a +\epsilon_{kl 3} A^a_l). \label{extraIntNA}\ee

Let us see now if~(\ref{extraIntNA}) affects perturbation theory. To this purpose note that the substitution in~(\ref{subL}) also shows that the term~(\ref{extraIntNA}) is  accompanied by another term with $\partial_\phi A_k^a +\epsilon_{kl 3} A^a_l$ substituted by $-\partial_t A_k^a$, which is the only interaction term involving the time-derivative of the gauge fields. One  can express both of them in terms of  $\mathscr{R}^{-1}(t\vec\Omega) A^a(x^\omega)$ and perform a change of variables (a rotation of angle $t\Omega$ around $\vec\Omega$, which has unit Jacobian)
  in the spatial integration,  in order to calculate in the perturbative expansion
  the
 correlators with field insertions of the form $O^\omega(x_1^\omega)...O^\omega(x_n^\omega)$. But doing so one finds that the $\Omega$ dependence in the interactions disappears there.  
So,  although $\Omega$ affects the propagators no $\Omega$-dependent vertices are introduced in perturbation theory, like in the general fermion-scalar theories of Refs.~\cite{Salvio:2025rma,Salvio:2025ggj}.

 The (gauge-dependent) interactions of the Faddeev-Popov ghosts with the other fields may also depend on $\Omega$. To illustrate this point let us consider the gauges~(\ref{tauGauge}) (including the $\xi_g$  gauges and thus the Feynman gauge). From~(\ref{GhostInttau}) one  sees that, expressing the interactions of the Faddeev-Popov ghosts in terms of \{$\mathscr{R}^{-1}(t\vec\Omega) A^a(x^\omega)$, $c^b(x^\omega)$, $\bar c^a(x^\omega)$\} 
and performing the above-mentioned change of variable in the spatial integration, again the $\Omega$ dependence in the interactions disappears. So in perturbation theory there are no $\Omega$-dependent vertices involving Faddeev-Popov ghosts either.

Finally, let us see  if/how rotation affects interactions in the matter sector of a gauge theory. As shown in Refs.~\cite{Salvio:2025rma,Salvio:2025ggj}, $\Omega$ appears only in the quadratic part of the matter Lagrangian of a pure scalar-fermion theory. However, introducing gauge fields, there is one $\Omega$-dependent interaction between gauge fields and matter, which appears in the scalar sector: this comes from the gauge kinetic terms of the scalars $\varphi$  in the Lagrangian density
\be \frac12 [(\partial_0-\Omega\partial_\phi +ig\theta^a A^a_0)\varphi](\partial_0-\Omega\partial_\phi +ig\theta^a A^a_0) \varphi,\label{gaugeKinS}\ee
where the $\theta^a$ are the generators of $\mathcal{G}$ in the representation of the scalars.
Again one can see that this $\Omega$-dependent interaction produces no $\Omega$-dependent vertices in perturbation theory. To  see this one can express~(\ref{gaugeKinS})   in terms of $\mathscr{R}^{-1}(t\vec\Omega) A^a(x^\omega)$ and $\varphi(x^\omega)$ and perform again the above-mentioned change of variable in the spatial integration.

So in perturbation theory, only the propagators are modified by $\vec\Omega$ and the $\mu_a$, the vertices remain unmodified in a general gauge theory with an arbitrary matter sector.  For the computation of the vertices one can then use well-known results from the literature (see e.g.~Refs.~\cite{Bellac:2011kqa} and~\cite{Landsman:1986uw}); on the other hand, the propagators have been computed, including the effect of $\vec\Omega$ and generic $\mu_a$,  in Refs.~\cite{Salvio:2025rma} (for scalars) and~\cite{Salvio:2025ggj} (for fermions) and in Sec.~\ref{propGauge} (for gauge fields and Faddeev-Popov ghosts). This provides a practical and model-independent recipe to perform perturbation theory in the presence of both $\vec\Omega$ and the $\mu_a$. 

For example, one can compute interaction rates (including decay and production rates) of a particle, which is weakly-coupled to the thermal plasma, with the Kobes-Semenoff rules~\cite{KS,KS2}, an extension to TFT of the well-known cutting rules in non-statistical quantum field theory. In~\cite{KS,KS2} Kobes and Semenoff assumed $\vec\Omega = 0$ and $\mu_a = 0$, but, as shown above, only the propagators are modified by $\vec\Omega$ and $\mu_a$, the vertices remain unmodified.  

\section{Summary and conclusions}\label{Conclusions}

Let us conclude by providing a summary of the main original results obtained.
\begin{itemize}
\item After a brief explanation of how to compute ensemble averages in the free spin-1 case and of the importance of using functional  and path-integral methods for gauge theories in Sec.~\ref{Preliminary remarks on free spin-1 particles}, the axial gauge was adopted to define all quantum gauge theories for a general equilibrium density matrix $\rho$ in Secs.~\ref{Classical gauge theory} and~\ref{Quantization and general path-integral formula} (including a generic matter sector).
\item Path-integral expressions for $Z$ and all  Green's functions of generic operators were also obtained in Sec.~\ref{Quantization and general path-integral formula} for any  $\rho$ at thermal equilibrium, first in the axial gauge and then in a generic gauge, which may or may not involve Faddeev-Popov ghosts.  The gauge invariance of the partition function and the Green's functions of gauge-invariant operators is clear from this construction.
\item The generalized KMS conditions for any pairs of operators $\mathcal{O}_1$ and $\mathcal{O}_2$ in arbitrary representations of the Lorentz and internal-symmetry group were derived in Sec.~\ref{Generalized Kubo-Martin-Schwinger condition} for all values of $\Omega$, $\mu_a$ and $T$.  An analyticity property of the non-time-ordered 2-point functions of $\mathcal{O}_1$ and $\mathcal{O}_2$ was established and allowed us to write the  generalized KMS conditions not only in coordinate space but also in momentum space.
\item In Sec.~\ref{propGauge} the generalized KMS conditions were applied to determine all needed thermal propagators of a gauge TFT for any $\rho$ at thermal equilibrium. 
\item The interactions of such a general gauge TFT were then studied in Sec.~\ref{Interactions}. Among other things,  that section established that, in perturbation theory, $\Omega$ (as well as the $\mu_a$) only  affects the thermal propagators, not the vertices. Since the thermal propagators were previously obtained in Sec.~\ref{propGauge}, this result provides us with a complete and general recipe to perform perturbation theory. 
\end{itemize}
Finally, let us remark that the model-independent approach of this paper guarantees that the results obtained can be used in the future in any specific gauge TFT model of physical interest at thermal equilibrium.
 \vspace{0.7cm}
 
 \subsection*{Acknowledgments}
I thank Francesco Tombesi for valuable discussions on rotating plasmas around black holes.

\vspace{1cm}

 \footnotesize
\begin{multicols}{2}

\end{multicols}

  \end{document}